# Energy and Eigenstate Using First Order Perturbation Theory

Raghu Ramachandran, Rama Komaragiri
National Institute of Technology, Calicut, India

October 30, 2009

**Abstract**

The energy and eigenstate of a perturbed system with a basis such that $|\psi> = \sum_j b_m |\varphi_m>$, are found by summation over the basis.

## 1 Introduction

Traditional descriptions of perturbation theory do not differentiate between the eigenstate of a system $|\psi\rangle$ and its basis functions $|\varphi_m\rangle$, such that $|\psi> = \sum_j b_m |\varphi_m>$ [1], [2].

If a system described by Hamiltonian $H$ is perturbed by $H'$, the energy and eigenstate for the system can be obtained by summation over the basis functions $|\varphi_m>$, as described below.

## 2 Energy and Eigenstate

Consider a system with an unperturbed Hamiltonian H, with an eigenstate $|\psi>$ and basis vectors $|\varphi_m>$, such that $|\psi> = \sum_j b_m |\varphi_m>$. The system is perturbed such that the Hamiltonian $H_1 = H + xH' + x^2 H''$ has an eigenstate $|\psi_1> = |\psi> + x|\psi>' + x^2 |\psi>''$. For the perturbed system

$$H_1 |\psi_1> = E_1 |\psi_1> \quad \text{eqn. (1)}$$

such that $E_1 = E + xE' + x^2 E''$. Upon substituting the above into eqn. (1), expanding and comparing coefficients one obtains

$$H|\psi> = E|\psi> \quad \text{eqn. (2)}$$
$$H'|\psi> + H|\psi>' = E'|\psi> + E|\psi>' \quad \text{eqn. (3) and}$$
$$H'|\psi>' + H|\psi>'' + H''|\psi> = E|\psi>'' + E'|\psi>' + E''|\psi> \quad \text{eqn. (4)}$$

Assuming $|\psi>' = \sum_j a_n |\varphi_n>$ and substituting into eqn. (3), $n \in \{j\}$

$$H'|\psi> + H \sum_j a_n |\varphi_n> = E'|\psi> + E \sum_j a_n |\varphi_n> \quad \text{eqn. (5)}$$

$$H'|\psi> + H\sum_j a_n|\varphi_n> = E'\sum_j b_n|\varphi_n> + E\sum_j a_n|\varphi_n> \quad \text{eqn. (6)}$$

Multiplying eqn. (6) by $<\varphi_m|, m \in \{j\}$

$$<\varphi_m|H'|\psi> + a_m E_m = E'b_m + Ea_m \quad \text{eqn. (7)}$$

$$E = <E> = \sum_j |b_m|^2 E_m$$

By superposition, from eqn. (3) $H_1|\varphi_{m1}> = E_1|\varphi_{m1}>$, and expanding results in

$$H'|\varphi_m> + H|\varphi_m>' = E'|\varphi_m> + E|\varphi_m>' \quad \text{eqn. (8)}$$

From eqn. (8),

$$H'|\varphi_m> + Hc|\varphi_m> = E'|\varphi_m> + Ec|\varphi_m> \quad \text{eqn. (9),}$$

here $|\varphi_m>' = c|\varphi_m>$

Multiply eqn. (9) by $<\varphi_n|$, $<\varphi_n|H'|\varphi_m> = E'<\varphi_n|\varphi_m>$

Therefore $E' = <\varphi_n|H'|\varphi_n>$

The $n^{th}$ perturbed energy level $E_{1n} = E_n + x<\varphi_n|H'|\varphi_n> \quad \text{eqn. (10)}$

Multiply eqn. (10) by $|b_n|^2$, $|b_n|^2 E_{1n} = |b_n|^2 E_n + |x<\varphi_n|H'|\varphi_n>$

$$\sum_j E_{1n}|b_n|^2 = \sum_j E_n|b_n|^2 + x\sum_j |b_n|^2 <\varphi_n|H'|\varphi_n>$$

$$E_1 = E + x\sum_j |b_n|^2 <\varphi_n|H'|\varphi_n>$$

where $E_1$ is the total energy of the perturbed system described by Hamiltonian $H'$.

From eqn. (7)

$$<\varphi_m|H'|\psi> - E'b_m = a_m(E - E_m)$$

$$\sum_j \frac{<\varphi_m|H'|\psi> - E'b_m}{E - E_m}|\varphi_m> = \sum_j a_m|\varphi_m>$$

$$\psi_1 = \psi + x\psi' = \psi + x\sum_j \frac{<\varphi_m|H'|\psi> - E'b_m}{E - E_m}|\varphi_m>$$

where $\psi_1$ is the eigenstate of the perturbed system described by Hamiltonian $H'$.

## 3 Summary

The total energy and the eigenstate of the system can be found by summation over the basis.